\documentclass[%
preprint,
showpacs,preprintnumbers,
 amsmath,amssymb,
 aps,
floatfix,
]{revtex4-1}

\usepackage{graphicx}
\usepackage{dcolumn}
\usepackage{bm}
\usepackage{alltt}
\usepackage{upquote}
\usepackage{color}
\usepackage{xspace}

\begin{document}


\title{The probability density function tail of the Kardar-Parisi-Zhang equation in the strongly non-linear regime}

\author{Johan Anderson$^1$}
\email{anderson.johan@gmail.com}
\author{Jonas Johansson$^2$}%
\affiliation{%
$^1$ Department of Earth and Space Sciences, Chalmers University of Technology, SE-412 96 G\"{o}teborg, Sweden\\
 $^2$ Solid State Physics and NanoLund, Lund University, Box 118, S-22100 Lund, Sweden}


\begin{abstract}
An analytical derivation of the probability density function (PDF) tail describing the strongly correlated interface growth governed by the nonlinear Kardar-Parisi-Zhang equation is provided. The PDF tail exactly coincides with a Tracy-Widom distribution i.e. a PDF tail proportional to $\exp( - c w_2^{3/2})$, where $w_2$ is the the width of the interface. The PDF tail is computed by the instanton method in the strongly non-linear regime within the Martin-Siggia-Rose framework using a careful treatment of the non-linear interactions. In addition, the effect of spatial dimensions on the PDF tail scaling is discussed. This gives a novel approach to understand the rightmost PDF tail of the interface width distribution and the analysis suggests that there is no upper critical dimension.
\end{abstract}

\maketitle

\section{Introduction}
In nature there are many important phenomena that are driven far from equilibrium by instabilities or by external forces. Examples are diverse from forest fires driven by the wind and impeded by moisture content to interstellar turbulence which is constantly stirred by supernova explosions. A proper description and understanding of the multiscale interactions that are responsible for the inevitably complex dynamics in these nonequilibrium systems remains a significant challenge in classical physics. 

The out of equilibrium interfacial growth is another example that has attracted much attention during recent years. A description of these growth processes that have been widely recognized is a Langevin like equation formulated by Kardar-Parisi-Zhang (KPZ) in a seminal paper see Ref. [\onlinecite{a10}]. The KPZ equation is one of the simplest non-linear generalizations of the diffusion equation and is thus connected to many other areas of non-equilibrium dynamics such as Burgers turbulence~\cite{a11, a12}, driven diffusion and dissipative transport~\cite{a122} as well as flame front propagation~\cite{a123}. The Burgers and KPZ is interconnected in such a way that the Burgers equation governs the dynamics of the local slope of the interface.

The KPZ equation has been studied extensively, however there are some remaining controversial issues, in particular the estimates of the upper critical dimension are in the range $d_c = 2.8-\infty$~\cite{a13, a131, HH2015, a132, a133, a134, a135, a136, a137, lassig1998, katzav2002, fogedby2008, canet2010, schwartz2012, alves2014}. Beyond $d_c$ the critical exponents are given by the mean field theory.

The purpose of the present work is to provide a statistical theory of interfacial growth in the strongly non-linear growth phase and thereby shed light on the elusive possibly finite upper critical dimension in the KPZ equation. We compute the tail of the probability density function (PDF) of the interface width using the instanton method in the Martin-Siggia-Rose (MSR) framework~\cite{a17}. The instanton method is a non-perturbative way of computing the PDF tails~\cite{thooft1, thooft2, a18, a19, a20, a16, cheklov1995, a21, a211, a22} taking all non-linear mode couplings into account. Historically, the instanton method was used in gauge field theory for calculating the transition amplitude from one vacuum to another vacuum~\cite{thooft1, thooft2}.

This is in contrast to other standard methods used to investigate scaling properties of non-linear Langevin equations such as the dynamic renormalization group (DRG) where an expansion in the non-linear term together with an average over the noise is employed. In DRG, counting the powers or the degree of the divergence usually pin-points to the critical dimension of the system. Furthermore, note that in the Edwards - Wilkinson model~\cite{EW1982} (found by suppressing the non-linear term in the Burgers equation) a different universality class is obtained whereas the Burgers and KPZ equations lie in the same class. This difference is stemming from a similarity in the nonlinear processes where interactions between different wave numbers are important, leading to a unique cascade, determining the scaling properties and the PDFs. From a theoretical point of view the PDF tails can be viewed as a transition amplitude from a quiescent state (where no growth occurs) to a final state determined by a coherent structure, from which the PDF tails are computed through a path-integral where the extremal path is determined by the instanton. Similarly to the Burgers equation the KPZ equation supports a coherent structure solution that can be used in the calculation as the path with highest probability and subsequently the path-integral can be solved using the saddle-point method. Without loss of generality, the PDF rightmost tail of a Langevin equation with a polynomial nonlinearity of finite grade is found by the instanton method in 1+1 dimension to be of the exponential form, $\exp{(- a \xi^s)}$ where $a$ is a positive constant, $\xi$ is the quantity described by the Langevin equation and $s$ can be determined as $s = (N+1)/m$. Here $N$ is the order of the non-linear term and $m$ is the sought moment~\cite{a22}. One particularly notable example here is the right tail of the PDF for velocity differences in Burgers equation that is found to have second order non-linear term ($N=2$) and first moment ($m=1$) and thus $s=3$~\cite{a19, a20}. This PDF tail was later corroborated by numerical simulations in Ref.~\cite{cheklov1995}. Other predictions of the instanton method have been shown to agree very well with numerical simulations, e.g. in a model of self-organization of sheared flows~\cite{a21} as well as that of heat flux~\cite{a211}. 

Regarding the Burgers equation and the KPZ equations, the PDF has been computed in a similar manner previously in Refs.~\cite{a138, a139, a1391, a1392, a15, a14, fogedby1998, fogedby2003, fogedby2009}, however all these results relies on the assumption of a weakly non-linear system. In the present setting we focus on the effects in the intermittent or strongly non-linear regime where the extremal solution dominates the behavior and scaling of the system in the long time limit. We find that the PDF exhibit heavier tails than a Gaussian distribution whereas the tails are subdominant to the exponential distribution. Under the assumption of isotropic growth in d-dimensions we find a smooth variation as the dimensionality $d$ is increased. This suggests that there exists no upper critical dimension and this is in accordance with the numerical work in Ref.~\cite{a13}. Furthermore it should be noted that this scaling of the PDF tail coincides with a Tracy-Widom distribution \cite{TW1, TW2}. 

The paper is organized as follows. First, the Kardar-Parisi-Zhang model is introduced and the PDF tail is computed using the instanton method. Then the result of the instanton PDF tails are corroborated in the one dimensional case through a derivation using the Fokker-Planck method. Finally the generalization to higher spatial dimensions are introduced and the paper is concluded by results and discussion.

\section{The PDF tail of the KPZ equation}
Lateral interface growth can be described by the Kardar-Parisi-Zhang (KPZ) model~\cite{a10}, that is the time evolution of the height $h$ is proportional to the square of the height gradient modified by diffusion and a stochastic forcing,  
\begin{eqnarray}
\frac{\partial h}{\partial t} = \lambda (\nabla h)^2 + \nu \nabla^2 h + f. \label{eq:1.1}
\end{eqnarray}
Here $f$ is a white noise forcing with a short correlation time modeled by the delta function as
\begin{eqnarray}
\langle f(x, t) f(x^{\prime}, t^{\prime}) \rangle & = & \frac{1}{\epsilon \sqrt{\pi}}\delta(t-t^{\prime})\exp \{-((x-x^{\prime}/\epsilon)^2\} \nonumber \\
 & = & \delta(t-t^{\prime}) \kappa_{\epsilon} (x - x^{\prime}), \label{eq:1.2}
\end{eqnarray}
and $\langle f \rangle = 0$. The angular brackets denote the average over the statistics of the forcing $f$.
The 1+1 dimensional KPZ equation is equivalent to the noisy Burgers equation by the relations
\begin{eqnarray}
u & = & -\nabla h, \label{eq:1.3}\\
h & = & -\int u dx. \label{eq:1.4}
\end{eqnarray}
Using the relations in Eq. (\ref{eq:1.3})-(\ref{eq:1.4}), we now find the noisy Burgers equation to be
\begin{eqnarray}
\frac{\partial u}{\partial t}  + 2 \lambda u \nabla u - \nu \nabla^2 u = -\nabla f. \label{eq:1.5}
\end{eqnarray}
The Burgers equation is known to support the ramp and shock-like stationary coherent solutions~\cite{a11}. We will adopt the ramp solution of the form $u \propto x$, using the relations between Burgers and KPZ we can find the solution for $h$ of the form $h = a x^2 + b$ where the time evolution of $h$ and the non-linear term both will be proportional to $x^2$. The constants $a$ and $b$ can be determined by inputting the trial function into KPZ yielding $a = \frac{1}{4 \lambda}$ and $b = \frac{\nu}{4 \lambda}$. Using the parabola solution as a coherent structure for the instanton method we will now compute the PDF tails.   

The MSR approach provides a many body description in the form of instanton excitations of the morphology of a growing interface. In this method an exact solution to the stationary field equation is required such as the one found above. Furthermore, the coefficients in the time evolution of the instanton will be dependent on the choice of structure. This is to elucidate on the machinery providing the scaling behavior and the PDFs~\cite{fogedby1998, fogedby2003, fogedby2009}. 

We compute the PDF tails of the mean square height fluctuations (the interface width) where we give special attention to the effect of dimensional scaling by using the instanton method. Here the mean square height fluctuations are defined as
\begin{eqnarray}
w_2 = \frac{1}{A_L} \sum_r (h(r,t)-\hat{h})^2 \label{eq:1.6}
\end{eqnarray}
where $A_L$ is the area of the substrate with characteristic linear dimension $L$, and $\hat{h} = \sum_r h(r,t)/A_L$ is the average height of the surface.
The PDF tails of $w_2$ are expressed in terms of a path-integral using the Gaussian statistics of the forcing. The optimum path is then associated with the creation of a shortlived coherent structure (among all possible paths - the instanton) and then the action is evaluated using the saddle-point method on the effective action. The saddle-point solution of the dynamical variable $h(x,t)$ of the form $h(x,t) = F(t) \phi(x)$ is called an instanton if $F(t) = 0$ at $t=-\infty$ and $F(t) \neq 0$ at $t=0$ as initial condition. Note that, the function $\phi(x)$ here represents the spatial form of the coherent structure. We will first consider the $d=1$ case and then generalize the found PDF to arbitrary dimension $d$. The probability density function of $w_2$ can be defined as
\begin{eqnarray}
P(w_2) =  \langle \delta(M(h) - w_2) \rangle = \int d \xi e^{i \xi w_2} I_{\xi}, \label{eq:1.7}
\end{eqnarray}
where 
\begin{eqnarray}
I_{\xi} = \langle \exp(-i \xi M(h)) \rangle. \label{eq:1.8}
\end{eqnarray}
Here $M(h)$ is the general expression for the $m$-th moment ($h^m$), however we will restrict this study to the second moment, c.f. the discussion in the introduction. Following Ref.~\cite{a18} the integrand can then be rewritten in the form of a path-integral as
\begin{eqnarray}
I_{\xi} = \int \mathcal{D} h \mathcal{D} \bar{h} e^{-S_{\xi}}, \label{eq:1.9}
\end{eqnarray}
where effective action $S_{\xi}$ of the KPZ equation is expressed as
\begin{eqnarray}
S_{\xi} & = & -i \int dx dt \bar{h} \left( \frac{\partial h}{\partial t} - \lambda (\nabla h)^2 - \nu \nabla^2 h \right) \nonumber \\
& + & \int dx dx^{\prime} dt \bar{h}(x,t) \kappa_{\epsilon}(x-x^{\prime}) \bar{h}(x^{\prime},t) \nonumber \\
& + & i \xi \int dx dt h^2 \delta(t).
\end{eqnarray}
Utilizing the instanton function $h(x,t) = F(t) \phi(x) $ the action $S_{\xi}$ can now be recast into,
\begin{eqnarray}
S_{\xi} & = & -i \int dt \mu \left( c_1 \dot{F} - c_0 \lambda F^2  - c_2 \nu F \right) \nonumber \\
& + & c_4 \int dt \mu^2 \nonumber \\
& + & i \xi \int dt c_3 F^2 \delta(t). \label{eq:2.1}
\end{eqnarray}
Here it is pertinent to note that the separation of variables in the instanton function is only appropriate in the vicinity of the used coherent structure since this gives the major contribution to the path-integral. The conjugate variable is denoted $\bar{h} = \mu(t) \bar{\phi}$ and we have used the following definitions of the constants,
\begin{eqnarray}
c_0 & = & \int dx \bar{\phi}(x) (\nabla \phi)^2(x) \label{eq:2.2}\\
c_1 & = & \int dx \bar{\phi}(x) \phi(x), \label{eq:2.3} \\
c_2 & = & \int dx \bar{\phi}(x) \nabla^2 \phi(x), \label{eq:2.4} \\
c_3 & = & \int dx \phi^2(x), \label{eq:2.5}\\
c_4 & \approx & \int dx dy \bar{\phi}(x) \bar{\phi}(y), \label{eq:2.6}
\end{eqnarray}
Note that the constant $c_4$ is evaluated for small values of $\epsilon$ in Eq. (\ref{eq:1.2}) and in the higher dimensional case $c_2$ will change with the dimension $d$ and $c_2 \neq c_0$. We compute the first variational derivatives to minimize $S_{\xi}$ with respect to $F$ and $\mu$ in order to find the path with highest probability identified by the instanton or the extremum of the action as, 
\begin{eqnarray}
\frac{\delta S_{\xi}}{ \delta \mu} & = & -i \left( c_1 \dot{F} - c_0 \lambda F^2 - \nu c_2 F \right) + 2 c_4 \mu = 0, \label{eq:2.7} \\
\frac{\delta S_{\xi}}{ \delta F} & = & -i \left( -c_1 \dot{\mu} - 2 c_0 \lambda F \mu - \nu c_2 \mu \right) + 2 i \xi c_3 F \delta(t) = 0. \label{eq:2.8} \nonumber \\
\end{eqnarray}
Eqs. (\ref{eq:2.7}) - (\ref{eq:2.8}) constitutes a dynamical system for the instanton time function $F$ and its conjugate $\mu$, see Ref. ~\cite{a22} for a more general discussion. We remind the reader that although $\mu$ appears to be simply a convenient mathematical tool, it does have a useful physical meaning that should be noted; it arises from the uncertainty in the value of $h$ due to stochastic forcing. That is, the dynamical system with a stochastic forcing should be extended to a larger space involving this conjugate variable, whereby $F$ and $\mu$ constitute a uncertainty relation. The instanton solution follows from a particular path out of all possible functional values of $F$ and $\mu$, which minimizes the
action S. Furthermore, it has the interesting physical meaning of mediating the forcing and the observable whose PDFs are sought. We proceed by solving these equations for $t<0$ and matching the solution at $t=0$. Note that the instanton solution $F$ rapidly grows at $t=0$ with increasing $\xi$ while it vanishes as $t \rightarrow -\infty$ and that the PDF is computed at time $t=0$. We start by computing an additional relation for the time evolution of the conjugate variable $\mu$ expressed in the real variable $F$ by taking the time derivative on Eq. (\ref{eq:2.7}),
\begin{eqnarray}
c_1 \ddot{F} - 2 c_0\lambda \dot{F} F - \nu c_2 \dot{F} = -2i c_4 \dot{\mu}. \label{eq:2.9}    
\end{eqnarray}
We now substitute $\mu$ and $\dot{\mu}$ in Eq. (\ref{eq:2.9}) by using Eqs. (\ref{eq:2.7}) and (\ref{eq:2.8}) yielding a second order non-linear differential equation for $F$,
\begin{eqnarray}
c_1^2 \ddot{F} = \nu^2 c_2^2 + 3 \nu c_0 c_2 \lambda F^2 + 2 c_0^2 \lambda^2 F^3. \label{eq:3.1}  
\end{eqnarray}
By setting $v = \dot{F}$ we find that we can rewrite the time derivative as $v (dv/dF)$ and we can now perform the first integration,
\begin{eqnarray}
c_1^2 v^2 & = & \nu^2 c_2^2 F^2 + 2 \nu c_0 c_2 \lambda F^3 + c_0^2 \lambda^2 F^4 \nonumber \\
& = & F^2(\nu c_2 + c_0 \lambda F)^2. \label{eq:3.2}
\end{eqnarray}
Below we will utilize the relation between $\dot{F}$ and $F$ that can be expressed as,
\begin{eqnarray}
c_1 \dot{F} = \pm F (\nu c_2 + c_0 \lambda F). \label{eq:3.3}
\end{eqnarray}
We can now easily determine the instanton time dependence from the separable differential Eq. (\ref{eq:3.3}),
\begin{eqnarray}
F(t) & = & \frac{\nu c_2 \exp(At)}{H_0 - c_0 \lambda \exp(At)}, \label{eq:3.4}\\
H_0 & = & \frac{\nu c_2 + c_0 \lambda F(0)}{F(0)}, \\ \label{eq:3.5}
A & = & \frac{\nu c_2}{c_1}. \label{eq:355}
\end{eqnarray}
Where we have used to positive sign in (\ref{eq:3.3}) in order to have vanishing instanton function at $t \rightarrow -\infty$. Note that the equation for $F(t)$ gives $F(0)$ at $t=0$ and that $F$ vanishes at $t \rightarrow -\infty $. Furthermore, it is interesting to note that the constant $A$ is dependent on the strength of the diffusive term through $c_2$. We now have to determine a value of $F$ at $t=0$ as a function of $\xi$. Thus we integrate Eq. (\ref{eq:2.8}) over $(-\epsilon, \epsilon)$ and use Eqs. (\ref{eq:2.7}) and (\ref{eq:3.3}) to obtain, 
\begin{eqnarray}
\mu(0) \approx 2i \frac{c_3 c_4}{c_0 c_1 \lambda} \xi. \label{eq:3.6}
\end{eqnarray}
Here it is assumed that $\mu$ is smooth and that the boundary condition $\mu(t>\infty) = 0$. The path-integral will now be computed using the saddle-point method in the limit of $\xi \rightarrow \infty$. First we have to evaluate the $\xi$-dependence of the action $S_{\xi}$. In the limit of $\xi \rightarrow \infty$, $S_{\xi}$ becomes
\begin{eqnarray} 
S_{\xi} & = & -i \int dt \mu \left( c_1 \dot{F} - c_0 \lambda F^2 - \nu c_2 F \right) \nonumber \\
& + & c_4 \int dt \mu^2 \nonumber \\
& + & i \xi \int dt c_3 F^2 \delta(t) \nonumber \\
& = & \frac{1}{4 c_4} \int dt (c_1 \dot{F} - c_0 \lambda F^2 - \nu c_2 F)^2 + i\xi c_3 F^2(0)\nonumber \\
& = & - \frac{c_1}{c_4} \int_0^{F(0)} dF (\nu c_2 F + c_0 \lambda F^2) + i \xi c_3 F^2(0) \nonumber \\
& = & - \frac{c_1}{c_4} \left( \nu c_2 \frac{F^2(0)}{2} + c_0 \lambda \frac{F^3(0)}{3}\right) + i \xi c_3 F^2(0) \nonumber \\
& \approx & \frac{8}{3} i \frac{c_3^3 c_4^2}{c_0^2 c_1^2 \lambda^2} \xi^3 - 4i \frac{c_3^3 c_4^2}{c_0^2 c_1^2 \lambda^2}\xi^3 \nonumber  \\
& = &  - \frac{4}{3}i \frac{c_3^3 c_4^2}{c_0^2 c_1^2 \lambda^2} \xi^3 \label{eq:3.7}
\end{eqnarray}
Here it is assumed that only the highest order term ($F(0)^3 \propto \xi^3$) contributes. Now let
\begin{eqnarray}  
\zeta = \frac{4}{3} \frac{c_3^3 c_4^2}{c_0^2 c_1^2 \lambda^2}, \label{eq:3.8}
\end{eqnarray} 
and the action becomes $S_{\xi} = -i \zeta \xi^3$. The tail of the PDF is then found by performing the $\xi$-integral in Eq. (\ref{eq:1.7}) by the saddle point method in the limit $w_2 \rightarrow \infty$. It is later shown that this corresponds to $\xi \rightarrow \infty$,
\begin{eqnarray}
P(w_2) & \sim & \int d \xi e^{i \xi w_2 - S_{\xi}} \label{eq:3.9}\\
& \approx & e^{i\xi w_2 + i \zeta \xi^3}. \label{eq:3.10}
\end{eqnarray}
We evaluate the $\xi$-integral using the extreme point $f^{\prime}(\xi_0) = 0$ and $\xi_0^2 = - w_2/(3\zeta)$ of $f(\xi) = i \xi w_2 + i \zeta \xi^3$ for the saddle-point method. This results in the PDF of $w_2$ as,
\begin{eqnarray}
P(w_2) \sim \exp\left(-\frac{2}{3} \frac{w_2^{3/2}}{\sqrt{3 \zeta}}\right) \label{eq:4.1}
\end{eqnarray}
where $\zeta$ is determined by Eq. (\ref{eq:3.8}).

\section{Computing the PDF tails using the Fokker-Planck equation}
In order to validate our found PDF we perform a similar analysis for $d=1$ where we compute the PDF using the Fokker-Planck (FP) method~\cite{a18, a22}. However, it seems to be highly non-trivial to generalize the PDFs found using the FP method to arbitrary spatial dimensions. Furthermore, the PDF is found to have the same fundamental exponential form whereas the coefficients differ.
We assume that we can write the functions ($h$ and $f$) in Eq. (\ref{eq:1.1}) as 
\begin{eqnarray}
h(x,t) & = & \phi(x) F(t), \label{eq:4.2}\\
f(x,t) & = & \phi(x) g(t), \label{eq:4.3}
\end{eqnarray}
with $\phi$ as before and $\langle g(t) g(t_1) \rangle = G \delta(t-t_1)$. Here $h(x,t)$ is the height function and $f(x,t)$ is the noise term in the KPZ system. Furthermore, we have assumed a similar separation of variables as in the previous instanton analysis based on the exact solution found.
By substituting this into Eq. (1) we find (neglecting the dissipation),
\begin{eqnarray}
\dot{F} = 4 \lambda a F^2 + g(t). \label{eq:4.4}
\end{eqnarray}
Furthermore we write the generating function ($z$) and the PDF ($P$) as
\begin{eqnarray}
z & = & e^{i \xi F}, \label{eq:4.5} \\
\langle z \rangle & = & \int dF P(F) e^{i \xi F} = \tilde{P} (\xi), \label{eq:4.6}\\
P(F) & = & \int d\xi e^{i \xi F} \langle z \rangle.  \label{eq:4.7}
\end{eqnarray}
Here the $\langle \cdot \rangle$ is the mean value integral over the forcing $f$. To compute the PDF from the FP equation we have to determine the time evolution of the generating function $z$ using Eq. (\ref{eq:4.5}),
\begin{eqnarray}
\frac{\partial z}{\partial t} & = & i \xi \dot{F} e^{i \xi F} \label{eq:4.8}\\
& = & i \xi (4 \lambda a F^2 + g(t)) z. \label{eq:4.9}
\end{eqnarray}
The time evolution of $P$ can now be found using Eq. (\ref{eq:4.5}) - (\ref{eq:4.7}) as
\begin{eqnarray}
\frac{\partial P}{\partial t} = - 4 \lambda a \frac{\partial}{\partial F}(F^2 P) - G \frac{\partial^2}{\partial F^2} P. \label{eq:5.1}
\end{eqnarray}
In the case of stationary PDF the differential equation is separable and we find
\begin{eqnarray}
P(F) = P_0 e^{- \frac{4 \lambda a}{G} F^3}, \label{eq:5.2}
\end{eqnarray}
where this can be rewritten in terms of the width function $w_2$ noticing that $F$ is supposed to be replaced by the second moment of the observable namely $w_2$, see e.g. \cite{a22}. Finally we find the PDF tails to be
\begin{eqnarray}
P(\hat{h}^2) = P_0 e^{- \frac{4 \lambda a}{G} (w_2)^{3/2}}. \label{eq:5.3}
\end{eqnarray}
This is the same stretched exponential dependence that was derived using the instanton method. We note that there are some differences in the constants of the exponential function. This is because somewhat different assumptions were used during the respective derivation of the PDF tails. In the FP method the dissipation was neglected while in the instanton method the value of the action was primarily dependent on the path in phase-space described by the instanton function. Hence differences in the constants arose.

\section{Dimensional scaling of the PDF tail}
In this section we will discuss one possible scaling of the PDF tail in higher spatial dimensions. One advantage of using the MSR path-integral methodology is that the PDF tails are governed by the instanton and thus determined by the non-linear interactions in the model whereas the physical form of the instanton (in 1D there are two solutions available: ramps and shocks) is of less importance. The spatial dependence determines the coefficients in the equation determining the time evolution of the instanton. In this paper we will model the effects of several spatial dimensions for the PDF tails found from the instanton method as being isotropic in all directions. Note that the KPZ has two terms involving the gradient operator where we assume $\nabla \phi(x_1,x_2, ... , x_d) \approx \vec{k} \phi(x_1, x_2, ... , x_d)$ where $k_d$ is the spatial length scale and $\phi$ is the spatial coherent structure in $d$ dimensions. Note that, in principle any coherent solution, such as the shock solution, can be used. However, that would entail in a change of the values of $c_0,...,c_4$. Here it is assumed that within the coherent structure we can approximate the gradient scale length to be $k_d$ and that it varies in the same manner in all directions. This is motivated by the fact that growth has no preferred direction. In Ref. \onlinecite{fogedby2009} simulations of the solutions to the KPZ are shown. In particular in Figure 9 in that investigation, height and slopes are shown. It is found that the height as a function of $x$ may be linearly growing or decaying as well as being approximately parabolic over finite distances. It is also shown that the slopes may be constant or linearly growing over finite distances. This indicates that solutions similar to our coherent structure indeed exist. Although, we neglect other possible solutions in higher spatial dimensions it is unlikely that these neglected structures would change anything beside the numerical coefficients in the equation and thus the same stretched exponential solution would be found. It is after all only the basic exponential scaling that is of interest in this work and it is most likely found by using the current assumption. There is a discussion of the actual spatial forms of the instanton in 2+1D and higher spatial dimensions in Ref.~\onlinecite{fogedby2009} where at least one isotropic form of the instanton is shown. In the isotropic case the terms become $\nabla^2 \phi = d k_d^2 \phi$ and $(\nabla \phi)^2 = d k_d^2 \phi^2 $. The generalization to the anisotropic case is straightforward, however, it does not add significantly to the understanding of scalings of dimensionality in the PDF tail. The coefficients in the found PDF transforms as,
\begin{eqnarray}
c_{0d} = d k_d^2 c_0, \label{eq:5.4} \\
c_{1d} = c_1, \label{eq:5.5}\\
c_{2d} = d k_d^2 c_2.  \label{eq:5.6}
\end{eqnarray}
Here the coefficients $c_{0d}, c_{1d}, c_{2d}$ represents the transformed coefficients in the $d$-dimensional case. We will show some results of the predicted PDF tails combining Eq. (\ref{eq:4.1}) with Eqs. (\ref{eq:5.4})-(\ref{eq:5.6}).

We will now elaborate on the effects of in particular dimensionality but also on the significance of the non-Gaussian tails by computing the dimensional dependence of the PDF tails. For the case of dimension $d=1$ we will compare the analytical prediction in Eq. (\ref{eq:4.1}) with an exponential tail, $P(w_2) \sim e^{-c d w_2}$, which we motivate below. The expression for the PDF tails can be written as 
\begin{eqnarray}
P(w_2) \sim e^{-c d w_2^{3/2}}, \label{eq:5.7}
\end{eqnarray}
where $c$ is a numerical coefficient and $d$ is the spatial dimension. Note that the power $3/2$ in the exponential comes directly from the non-linear interaction term in the KPZ equation whereas the factor $c d$ arises from the linear part of the equation. That is, there is no obvious relation to the standard scaling exponents. This suggests that KPZ supports harmonic growth in all spatial directions and the physical dimensions are weakly correlated with a smooth behaviour of the scaling function as the dimensionality changes. Furthermore, we note that one indication of an upper critical dimension may be a discontinuous change of the scale lengths, $k_d$ in Eqs. (\ref{eq:5.4}) and (\ref{eq:5.6}), as the spatial dimensionality, $d$, is changed. Here we emphasize that the PDF tails is only one way of characterizing the roughness of a surface. For example, an increase in the fourth moment (kurtosis) indicates a higher likelihood of extreme events, which may in turn indicate an increased roughness.

Much of the knowledge about the KPZ equation come from simulations on systems, typically lattice models, that belong to the KPZ universality class. The problem with such lattice models is that given a specific model, such as the restricted solid-on-solid model~\cite{a23}, the KPZ parameters, $\lambda$ and $\nu$ are fixed and the non-linear parameter, $\lambda$, is often quite small~\cite{a24}, which result in PDF tails that are indistinguishable from exponentially decaying PDF tails~\cite{a1391} or result in stretched exponentials with exponent slightly smaller than one~\cite{a26}. This is is contrast to our exponent of 3/2, which holds in the strongly non-linear regime or strongly correlated interfacial growth. Due to numerical instabilities, which seem to grow with increasing $\lambda$, it is a very challenging task to solve the KPZ equation by numerical integration. It has been proposed to use an exponential to impede numerical instabilities however using the instanton method this form would give rise to a different scaling of Gumbel type ~\cite{anderson2010}. Thus, we are not aware of any numerical estimations of the solutions to the KPZ equation in the strongly non-linear regime with which we could compare our analytical results.

\section{Results and discussion}
In finding the analytical solution for the PDF tail we assumed that we solved for values in the rightmost part of the tail due to taking $\xi \rightarrow \infty$ and using the saddle-point method whereas here we will show the PDF for all values of $w_2/\langle w_2 \rangle > 1$. Here it is pertinent to stress that the analytical method are only valid for the tail of the PDF. The constant $c$ in Eq. (\ref{eq:5.7}) includes an unknown forcing strength and can be estimated from simulations, e.g. those reported in Ref.~\cite{a13}.
\begin{figure}
  \includegraphics[height=.3\textheight, width = 8.5cm]{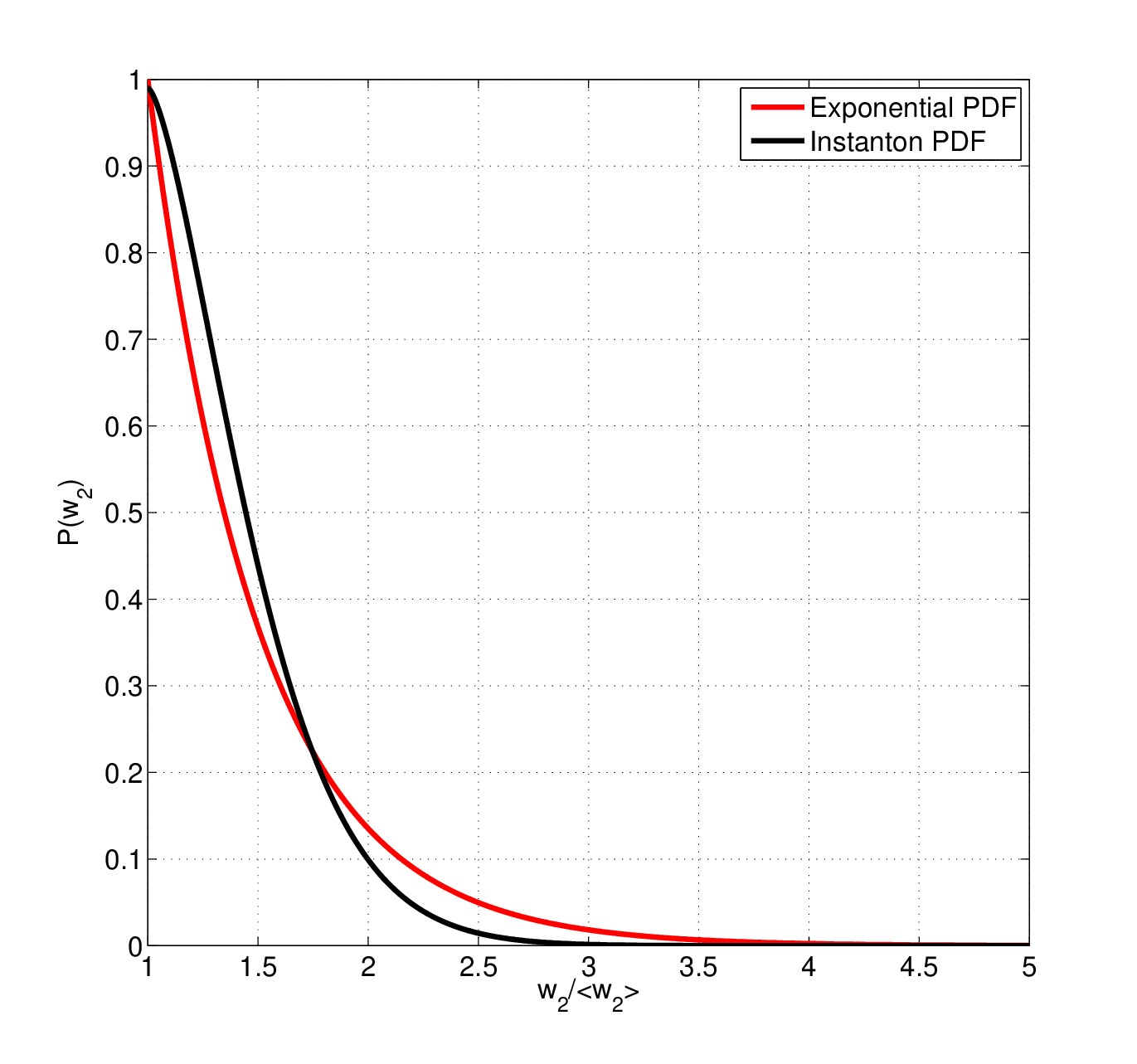}
  \caption{(Color online). The scaling function of the width distribution $P(w_2)$ as a function of $w_2/\langle w_2 \rangle$ (linear scale) where the results are shown for exponential PDF (red) and instanton PDF (black). The value of the constant was $c = 0.86$.}
\end{figure}
In Figure 1 we show the scaling function of the width distribution $P(w_2)$ as a function of $w_2/ \langle w_2 \rangle$ where quite small differences between the two distribution tails are present. Note that the PDF is normalized in such a way that the total probability is unity and that $P(1) \approx 1$ whereas the values of $w_2/\langle w_2 \rangle$ are arbitrary. The maximum value of the PDF is determined by the strength of the forcing.
Figure 2 displays again the scaling function $P(w_2)$ as a function of $w_2/ \langle w_2 \rangle$ however this time in lin-log scale which highlights the significant differences for small probabilities.
\begin{figure}
  \includegraphics[height=.3\textheight, width = 8.5cm]{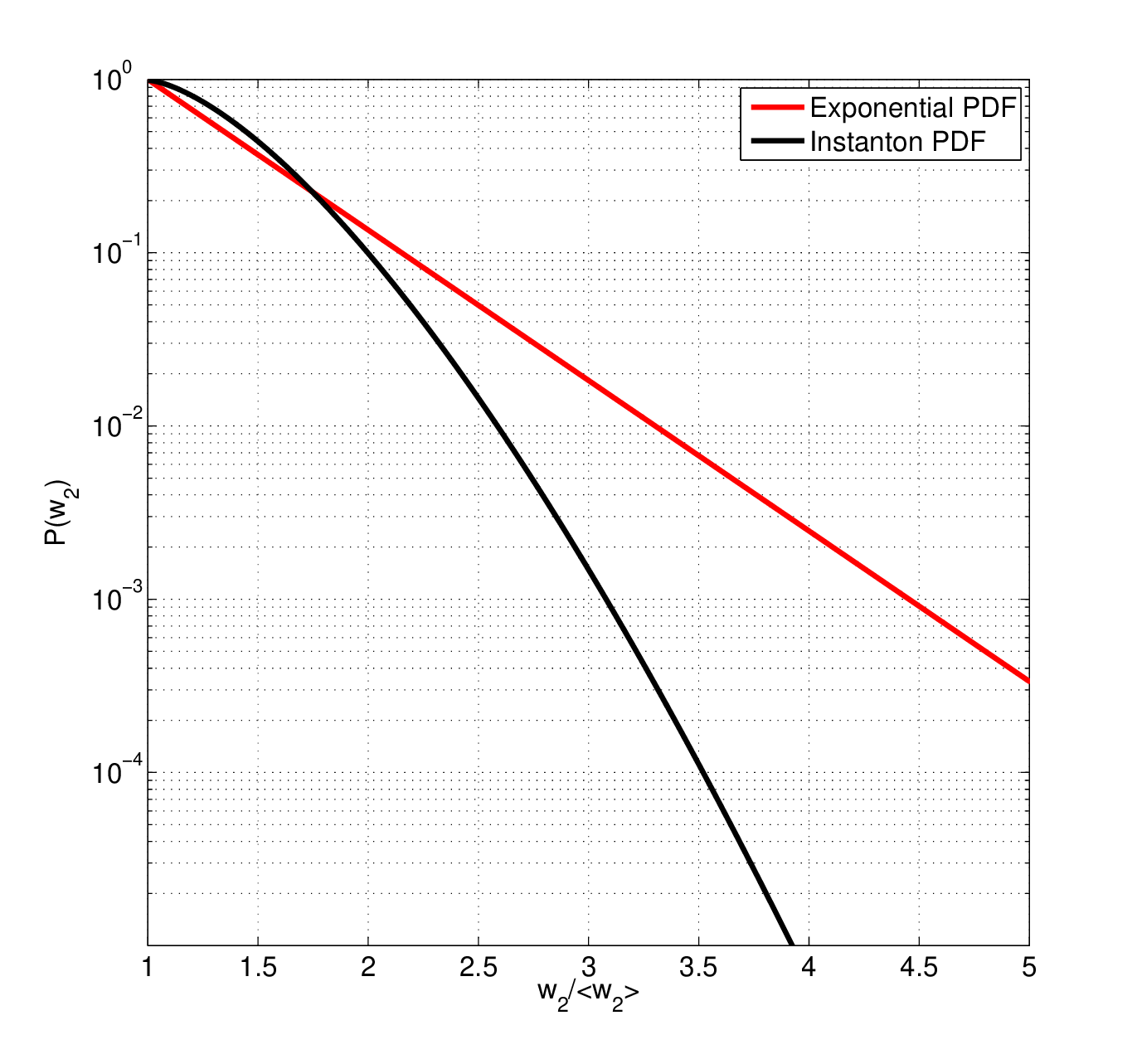}
  \caption{(Color online). The scaling function of the width distribution $w_2 P(w_2)$ as a function of $w_2/ \langle w_2 \rangle$ (log-linear scale) where the results are displayed for exponential PDF (red) and instanton PDF (black). The value of the constant was $c=0.86$.}
\end{figure}

We note that, although the PDF tail does not coincide with the plain exponential distribution we have found a salient tail that predicts the long time limit of strongly intermittent or strongly correlated interfacial growth. Here it is pertinent to remind the reader that 1+1D is a very special case and, as regards the steady-state statistics, the KPZ non-linearity is irrelevant in this dimension~\cite{HH2015} and to stress that the present method yields the exponential tail for the linear KPZ system in any dimension. More precisely the non-linear part is not important for the 1+1D PDF in steady-state but not all the properties are indeed independent of this non-linear term. For example, $\alpha$, the roughness exponent is indeed the same as in the linear Edwards-Wilkinson equation, but $z$, the dynamic exponent, is $z=3/2$ for KPZ and $z=2$ for Edwards-Wilkinson.

In particular, the solution presented here of the PDF tails has the same qualitative behaviour as was found in Ref.~\cite{a13} for increasing spatial dimensionality, i.e. we find a PDF tail that decreases faster with higher dimensions where no obvious higher critical dimension is visible.  However in 2+1D and up the situation is still not well understood. One complication with the present method is to find the proper higher dimensional structure in the derivation of the instanton. A promising and not fully explored numerical procedure to find the extremal paths in higher dimensions such as 2+1D and up can be found in Ref. \onlinecite{fogedby2009} however the generic scaling of the PDF in higher dimensions is not clear as is indicated in Ref. \onlinecite{HH2015}. The instanton method would in principle allow for a solution to the KPZ equation formulated in hyperspherical coordinates. As the appropriate solution in higher dimension is not precisely known, this approach is beyond the scope of the present work. Furthermore, the regular steady-state solution in dimension $d>1$ may coincide with the solution found here in the strongly correlated state or strongly non-linear regime. We would also like to remind the reader that the long time PDF found here is only one admissible solution generated for a particular coherent structure. However there might be many solutions to the KPZ in higher dimensions and thus multiple solutions are possible.

In addition, in Ref. \onlinecite{HH2015}, a suggestive parabolic relation between the skewness and the kurtosis for increasing spatial dimensions is shown. A parabolic relation between the third and fourth moment is a signature and common denominator of a system described by an equation including a quadratic nonlinearity \cite{sattin2009, krommes2008}. This lends support to the current modelling effort as this indicates that in higher dimensions the nonlinearity will be more prominent and determine the dynamics. Thus, in higher spatial dimension a deviation from a purely exponential PDF tail as the one reported for the 1+1D case~\cite{a1391} is strongly expected.

It is interesting to note that two recent papers using similar methods exploring large deviations in interfacial growth by the KPZ equation find the stretched exponential $\exp(-a w_2^{3/2})$ which coincides with the findings in this paper~\cite{merson2015, doussal2016}.  Moreover we note that this PDF also coincides with the Tracy-Widom distribution~\cite{TW1, TW2}. Note that, in Ref. \onlinecite{doussal2016} the appearance of the Tracy-Widom distribution is, within Random Matrix Theory, attributed to a third order phase transition between strong and weak coupling regimes.

To this end, in this paper we have computed the PDF tails in the KPZ model and evaluated the dependence of spatial dimensionality. The PDF tails of the form $\sim \exp( - c w_2^{3/2})$ (where $c$ is a numerical coefficient dependent on the dimension $d$) are relevant for intermittent events in the strongly non-linear growth regime. Moreover, the PDF tails are significantly alleviated compared to a Gaussian distribution whereas the tails are subdominant to the exponential distribution. Of particular interest are the effects of higher dimensions on the coefficient $c$. Here it is indicated that by using an isotropic growth model the coefficient $c$ changes smoothly as $d$ increases and thus suggests that there may not be an upper critical dimension $d_c$, in agreement with the discussion in Ref. \onlinecite{a13}.

\acknowledgments
Johansson wants to acknowledge NanoLund (the Center for
Nanoscience at Lund University) and the Swedish Research Council (VR) for financial support. Anderson acknowledges the Max-Planck Institute for its hospitality where most of this work was carried out.

\end{document}